\documentclass[prl, twocolumn,preprintnumbers,amsmath,amssymb]{revtex4}

\usepackage{graphicx}
\usepackage{dcolumn}
\usepackage{bm}
\usepackage{amssymb}
\usepackage{wasysym}
 \usepackage{float}

\usepackage{tabularx}
    \newcolumntype{Y}{>{\centering\arraybackslash}X}

\begin{document}

\preprint{}

\title{Strain gradient induced polarization in SrTiO$_3$ single crystals}

\author{P. Zubko}
\email{pz212@cam.ac.uk}
\author{G. Catalan}
\email{gcat05@esc.cam.ac.uk}
\author{A. Buckley}
\author{P. R. L. Welche}
\author{J. F. Scott}
\affiliation{%
Centre for Ferroics, Department of Earth Sciences, University of Cambridge, Cambridge CB2 3EQ, United Kingdom 
}%

\date{\today}

\begin{abstract}
Piezoelectricity is  inherent only in noncentrosymmetric materials, but a piezoelectric response can also be obtained in centrosymmetric crystals if subjected to inhomogeneous deformation. This phenomenon, known as flexoelectricity,  affects the functional properties of insulators, particularly thin films of high permittivity materials. We have measured strain-gradient-induced polarization in single crystals of paraelectric SrTiO$_3$ as a function of temperature and orientation down to and below the 105 K phase transition. Estimates were obtained for all the components of the flexoelectric tensor, and calculations based on these indicate that local polarization around defects in SrTiO$_3$ may exceed the largest ferroelectric polarizations. A sign reversal of the flexoelectric response  detected below the phase transition suggests that the ferroelastic domain walls of SrTiO$_3$ may be polar.

\end{abstract}

\maketitle

Flexoelectricity is the coupling between dielectric polarization and strain gradient. Because strain gradients break inversion symmetry, flexoelectricity allows extracting charge from deformations even in materials that are not piezoelectric. 
The historical development of this phenomenon has been summarized in the reviews by Tagantsev \cite{TagantsevReview} and more recently by Cross \cite{CrossReview} and Sharma \emph{et al.} \cite{Sharma}.
For decades after its proposal in the early 1960s \cite{MashkevichTolpygo, Kogan}, there were very few studies of flexoelectricity, which may seem surprising given that it is a more general property than piezoelectricity. The reason  is  the comparatively small magnitude of the flexoelectric coefficient $f\sim e/a$ (where $f$ is the constant of proportionality between polarization and strain gradient, $e$ the electronic charge and $a$ the lattice parameter) in typical dielectrics which, combined with the small size of strain gradients in bulk samples, means that the flexoelectric effect is generally minuscule.  Two recent developments, however, are changing the perceived importance of flexoelectricity. First, a predicted proportionality between flexoelectric coefficient and dielectric constant \cite{Kogan, TagantsevPRB} has inspired Cross and coworkers  to conduct a series of seminal experiments in which they have measured the flexoelectric coefficients of high permittivity materials such as relaxors and ferroelectrics \cite{CrossReview, Ma_and_Cross}, and found them to be very large indeed. In addition, Fousek \emph{et al.} \cite{Fousek} realized that this large flexoelectric response could be used to produce piezoelectric composites from centrosymmetric materials by simply tailoring their shapes, an idea which was recently demostrated experimentally  by Zhu \emph{et al.} \cite{Zhu_and_Cross}. Second, recent works by Catalan \emph{et al.} \cite{Catalan} have shown that ferroelectric thin films can sustain large strain gradients due to lattice mismatch with the substrate, and that the flexoelectric effect of such gradients can have an important effect on their functional properties, particularly the frequently discussed lowering of permittivity in ferroelectric thin films.

In spite of its growing relevance, our understanding of flexoelectricity is still rather incomplete. On a theoretical level, there are still no first-principles studies. On an experimental level, only ceramic samples have been analyzed, with the problem that grain boundaries can contribute to measured charge due to their possibly polar nature \cite{Petzelt}, through tribological effects or via  surface piezoelectricity \cite{TagantsevReview}. Also, because of the need for a high dielectric constant, only materials with ferroelectric phases (or polar nanoregions in the case of relaxors) have been studied so far. This has the drawback for interpretation that ferroelectricity (and thus piezoelectricity) may persist above the nominal phase transition temperature due to local strain effects, and  contribute piezoelectrically to the measured charge.

We have chosen to work instead with single crystals of the paraelectric strontium titanate. At room temperature SrTiO$_3$ (STO) has the cubic perovskite structure and remains centrosymmetric even in the tetragonal phase below ~105 K. The absence of  piezo- and ferroelectric contributions, combined with relatively high dielectric permittivity, makes STO a natural choice for studying flexoelectricity. The use of single crystals of different orientations also allows all the flexoelectric tensor components of STO to be determined.

High purity STO single crystals of various dimensions (typically 3--5mm wide, 5--15mm long and 50--500$\mu$m thick) were obtained from PiKem Ltd. Impurity levels (in parts per million) determined by the supplier are:  Ni=3, Fe=2, Cr$<$2, Ba, Na and Si each $<$1. Ca impurities at $<$1ppm are several orders of magnitude below the critical concentration of Ca$^{2+}$ ions above which a ferroelectric phase transition can be induced at low temperatures \cite{Bednorz}. The transparent colourless crystals were supplied with a surface  roughness of a few \r{A}.  Top and bottom Au electrodes of area 10--30mm$^2$ (depending on crystal size) were deposited by sputtering. Pt wires (50$\mu$m in diameter) were attached to the Au electrodes with silver paste which was then annealed at 130$^\circ$C to improve conductivity and mechanical robustness. 

\begin{figure}[h]
\centering
\scalebox{0.35}{\includegraphics{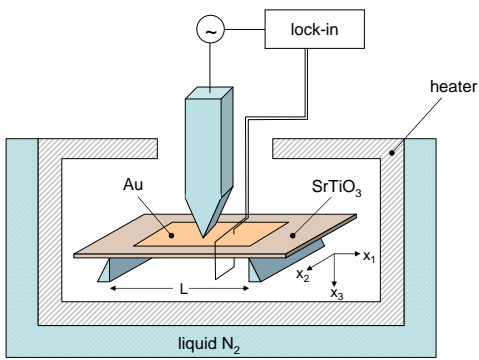}}
\caption{Experimental setup for flexoelectric measurements.}
\label{ExptSetup:OK}
\end{figure}

The experimental setup is sketched in figure~\ref{ExptSetup:OK}. A dynamical mechanical analyzer (DMA) with an insulating quartz probe was used to induce oscillatory bending (typically driven at 30-40Hz) and measure its amplitude. A static stress was applied simultaneously to hold the sample in place. For temperature dependent measurements, heating and cooling could be achieved by competitive action of a resistive heater and a liquid N$_2$ bath. The displacement current $I$ due to the induced polarization was measured using a Signal Recovery 7265 dual phase lock-in amplifier. The maximum strains (approx. 10$^{-6}$) achieved by the bending during the experiments were far below those which, according to thermodynamic \cite{Pertsev, Haeni} and \emph{ab-initio} \cite{Lin} calculations, are capable of inducing a polar phase in the investigated temperature range. The crystals are therefore expected to be neither ferro- nor piezoelectric. A direct measurement of piezoelectricity showed no signal above the noise level placing an upper limit on the piezoelectric coefficient $d_{33}$ of 0.03pC/N.   In addition, no signs of second harmonic generation (SHG) could be detected.

\begin{figure}[h]
\centering
\scalebox{0.76}{\includegraphics{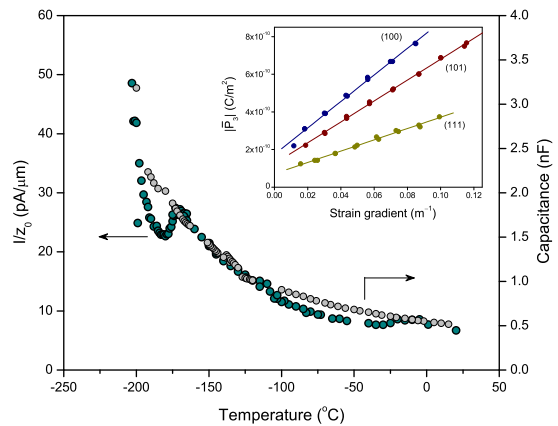}}
\caption{Temperature dependence of the ratio $I/z_0$ (which is proportional to $\mu$) and of the  capacitance for a sample polished down to 100$\mu$m using diamond impregnated lapping film. The flexoelectric response increases with the dielectric constant, showing an anomaly at the ferroelastic phase transition. The absolute temperature values shown may be a few degrees too low due to some thermal lag between the thermocouple and the sample. Inset shows the linear dependence of flexoelectric polarization on strain gradient  for 300$\mu$m thick STO crystals of  different orientations at room temperature.}
\label{temp_dep:OK}
\end{figure}

The strain gradient was derived from the usual equation for a bent beam (e.g. see \cite{Landau_Lifshitz}) as 
\begin{equation}
\frac{\partial \epsilon_{11}}{\partial x_3}=3z_0\left(\frac{L}{2}\right)^{-3}\left(\frac{L}{2}-x_1\right),
\end{equation}
where  $L$ is the distance between the knife edges (in our case $L=$10, 7.5,  or 5mm), $z_0$ is the displacement at the centre as measured by the DMA and the distances $x_i$ are measured from the centre of the crystal. 

Flexoelectricity is described by a fourth rank tensor $f_{ijkl}$:
\begin{equation}
P_i=f_{ijkl}\frac{\partial \epsilon_{kl}}{\partial x_j},
\label{definition_of_f:OK}
\end{equation} 
where $P$ is the flexoelectric polarization and $\epsilon_{kl}$ is the symmetrized elastic strain tensor. If $\omega/2\pi$ is the frequency of the applied mechanical stress and $A$ is the electrode area, the average out of plane polarization can be computed by measuring the ac current produced by bending and using  $\overline{P}_3=I/\omega A$. The effective flexoelectric coefficient $\mu$ was calculated from the measured average polarization and strain gradient using
\begin{equation}
\overline{P_3}=\mu\overline{\frac{\partial \epsilon_{11}}{\partial x_3}} \textrm{\ \ \ \ \ and \ \ \ \ } \overline{\frac{\partial\epsilon_{11}}{\partial x_3}}=\frac{12z_0}{L^3}(L-a), 
\end{equation}
where $a$ is the half-length of the electrodes.

From phenomenological arguments, the flexoelectric coefficient is expected to be proportional to the dielectric constant \cite{Kogan, TagantsevPRB}. To test this, the flexoelectric current was measured as a function of temperature   (figure~\ref{temp_dep:OK}). As expected, the current increases upon cooling, qualitatively following the dielectric permittivity. On approaching the ferroelastic transition, however, an anomaly, not present in the dielectric constant, is seen in the flexoelectric response. 
\begin{figure}[h]
\centering
\scalebox{0.33}{\includegraphics{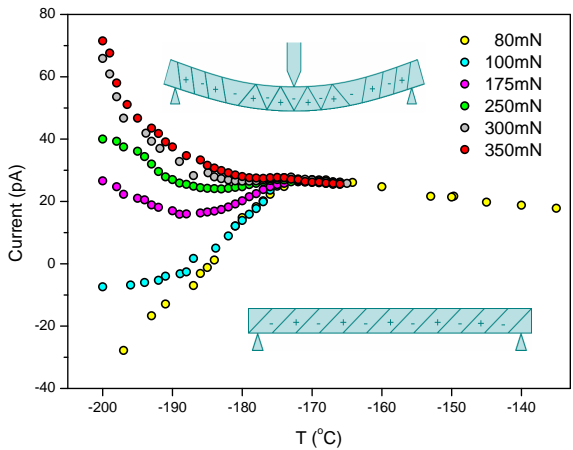}}
\caption{Temperature and static force dependence of the flexoelectric current for the 100$\mu$m sample in figure~\ref{temp_dep:OK}.
As the static force is increased, the domain walls become  less mobile due to impingement and a recovery of the positive flexoelectric current is observed.
Similar behaviour was observed for the original thicker crystals.
Domains with the c-axes along \textbf{x}$_1$ and \textbf{x}$_3$ are labeled $+$ and $-$ respectively.}
\label{TempDep:OK}
\end{figure}

Previous studies \cite{Buckley,Kityk} have shown that ferroelastic domains appear below ~105K and are responsible for the large softening of the elastic modulus in STO. The relaxation of strain gradients due to domain wall readjustments may therefore be expected to reduce the flexoelectric response, as indeed observed. This gradient relaxation by domain readjustment can be studied by changing the static force applied to the crystal. With large static force, the crystal is very bent and the domain walls impinge on each other, so that they can no longer move under the dynamic load. Thus, the mechanical and flexoelectric response should approach that of a monodomain crystal. For low static forces, the domain walls are quite free to move and thus the relaxation of the strain gradient is maximum. Accordingly, one would expect the flexoelectric current to drop. Instead, however, we observed the current to drop through zero to a \emph{negative} value (bottom curves in figure~\ref{TempDep:OK}). 
The fact that the change of sign in the current is only seen when domain walls can move leads us to believe that the domain walls may be charged with a polarization of opposite sign to that of bulk flexoelectricity, so that their motion under the periodic stress produces a current of the opposite sign to the flexoelectric response. 

The possible polarization of domain walls in STO (a non-polar material) is an unexpected result. At present, we can think of three explanations for the observed domain wall charge: (i) the local strain gradient at the walls polarizes them through flexoelectricity, (ii) order-parameter coupling between the ferroelastic distortion and the (suppressed) ferroelectric polarization induces polarization at the domain wall \cite{Salje}, or (iii) the domain walls trap charged deffects such as oxygen vacancies \cite{Vanderbilt}. 

\begin{table}[h] 
\begin{tabularx}{\linewidth}{
>{\setlength{\hsize}{1.0\hsize}}Y
>{\setlength{\hsize}{1.0\hsize}}Y
>{\setlength{\hsize}{1.0\hsize}}Y
>{\setlength{\hsize}{1.0\hsize}}Y}
\hline
$\textrm{\textbf{x}}_1$ & $\textrm{\textbf{x}}_2$ & $\textrm{\textbf{x}}_3$ & $\mu$ (C/m) \\
\hline

[100]	&	[010]	&	[001]	&	$+6.1\times 10^{-9}$\\

[10\=1]	&	[010]	&	[101]	&	$-5.1\times 10^{-9}$\\

[11\=2]	&	[\=110]	&	[111]	&	$-2.4\times 10^{-9}$\\
\hline

\end{tabularx} 
\caption{Orientation dependence of the flexoelectric response.} \label{Orientation:OK}
\end{table}
We have also measured the flexoelectric response of [001], [101] and [111]-oriented samples (inset of figure \ref{temp_dep:OK}). The corresponding in-plane orientations were determined by X-ray diffration and are shown in table~\ref{Orientation:OK}, together with the average measured flexoelectric coefficients. All the coefficients were found to be of the same order of magnitude (ranging between about 1 and 10nC/m),  but have different signs. While there was some inter-sample variation in the magnitudes of $\mu$ for each of the orientations, the signs were robust. 

For any material belonging to one of the cubic point groups there are only five independent components of the flexoelectric tensor $f_{ijkl}$. In the case of STO, which belongs to the $O_h$ group, the 4-fold rotation symmetry further reduces this number to three: $f_{1111}$,  $f_{1122}$ and $f_{1221}$($=f_{1212}$). For different crystal orientations, the measured polarization arises from different combinations of the three flexoelectric tensor components, i.e., the calculated values of $\mu$ are effective coefficients rather than the flexoelectric tensor components defined in (\ref{definition_of_f:OK}). 
In addition we must not forget the contributions to $P_3$
from the gradients of $\epsilon_{22}$ and $\epsilon_{33}$. For a bent plate \cite{Landau_Lifshitz} $\epsilon_{33}=-\frac{c_{31}}{c_{33}}\epsilon_{11}
-\frac{c_{32}}{c_{33}}\epsilon_{22}$ where $c_{ij}$ are the elastic moduli. In our bending geometry, the anticlastic strain $\epsilon_{22}$ is assumed to be negligible, giving for [001] oriented samples
\begin{equation}
\mu=f_{3311}-\frac{\nu_{21}}{1-\nu_{21}}f_{3333}.
\label{mu:OK}
\end{equation}
The relevant anisotropic Poisson ratios $\nu_{ij}$ can be obtained from the known elastic moduli of STO \cite{Poindexter}.
For samples whose edges are not aligned with the crystallographic axes $\hat{\textrm{\textbf{e}}}_i$ the $f_{ijkl}$ above should be replced by 
$f'_{ijkl}=L_{ip}L_{jq}L_{kr}L_{ls}f_{pqrs}$  with $L_{ij}=\hat{\textrm{\textbf{x}}}_i \cdotp \hat{\textrm{\textbf{e}}}_j$ and $\nu_{ij}$ by the corresponding $\nu '_{ij}$.

Inserting the values for $\mu$ and $\nu '_{ij}$ for the different orientations of the crystals and the corresponding expressions for $f_{ijkl}'$ in terms of $f_{ijkl}$ into equation (\ref{mu:OK}) leads to three simultaneous equations. However, it turns out that these are not independent and hence cannot be solved to find the individual tensor components, leaving us instead with the following relations:
\begin{eqnarray}
f_{1122}-\alpha f_{1111}&=&\mu_{\textrm{(001)}}\nonumber\\
\frac{1}{2}(1-\beta)(f_{1111}+f_{1122})-(1&+&\beta)f_{1212}=\mu_{\textrm{(101)}}\nonumber
\end{eqnarray} 
where $\alpha=\frac{c_{12}}{c_{11}}$ and $\beta={\frac{c_{11}+c_{12}-2c_{44}}{c_{11}+c_{12}+2c_{44}}}$. Nevertheless, the interdependence of the equations can be used to check the self consistency of  our analysis since we can use the measured $\mu$ values for the [001] and [101] to calculate the expected $\mu$ for [111] which turns out to be $-$3.1nC/m and compares reasonably well, considering the intersample variation, with the actual experimental value of $-$2.4nC/m. 
A third independent equation can be obtained by changing the geometry of the sample from a plate to a beam. 
In this case $\epsilon_{22}=-\nu_{21}\epsilon_{11}$  and $\epsilon_{33}=-\nu_{31}\epsilon_{11}$ \cite{Landau_Lifshitz}. Using this,  rough estimates of $f_{1111}\approx -9$nC/m, $f_{1122}\approx 4$nC/m and $f_{1212}\approx 3$nC/m were obtained \cite{rough}.

Taking these as order of magnitude values, we can estimate the expected flexoelectric polarization around defects such as dislocations in epitaxial STO.  Chu \emph{et al.} and Nagarajan \emph{et al.} \cite{dislocations} report strains at dislocations of order 0.05 relaxing over several nm thus giving rise to strain gradients of order 10$^7$m$^{-1}$. Flexoelectric coefficients of order $10^{-9}$--$10^{-8}$C/m will therefore give local polarizations of about 1--10$\mu$C/cm$^2$ at room temperature, and up to 100 times more at low temperatures due to the increase in dielectric permittivity. This implies that the local polarization around defects in non-polar materials such as STO can be bigger than the ferroelectric polarization in the best ferroelectrics. Obviously, this is only a local effect, but given high enough density of dislocations (as can happen in strained thin films) we should expect the impact of flexoelectric polarization on the functional properties of dielectrics to be very large.

Before concluding we should briefly address the possibility of artefacts. For high purity STO we do not expect any bulk piezoelectric contribution to the measured current. The lowering of the symmetry at the surface however, introduces the possibility of a contribution from surface piezoelectricity \cite{TagantsevPRB}. Experimental data and first principles calculations  suggest that the perturbed layer in only a few lattice constants thick, with no consensus as to whether it is polar or not \cite{Herger,Padilla_Vanderbilt}. 
Our estimates show that to mimic the measured response the piezoelectric modulus of the surface has to be approximately the same as that of a good ferroelectric such as BaTiO$_3$ which seems unlikely.
 Finally, whatever the nature of the surface, there will in general always be a surface flexoelectric contribution $f_{sf}\sim e/a$ \cite{TagantsevPRB}. However unlike the bulk effect,  $f_{sf}$ is not expected to scale with the dielectric permittivity \cite{Ma_and_Cross,TagantsevPRB} and thus it  should be  about two orders of magnitude lower than the bulk effect in STO. Nonetheless, recent works  have revealed extended near surface skin regions up to 100$\mu$m deep with local fluctuations of the ferroelastic phase transition \cite{Shirane}. Such regions were found to be inhomogeneously strained and thus may possess flexoelectric polarization even in the absence of external stress. This may play some role in our samples, but even more so in fine-grained ceramics due to their higher density of surfaces, which may contribute to the very high values of the flexoelectric coefficient obtained in ferroelectric ceramics. \cite{Ma_and_Cross}. 

To conclude, we have measured the dielectric polarization induced by bending in single crystals of SrTiO$_3$. Measurements of samples with different crystallographic orientations have allowed all components of the  flexoelectric tensor to be estimated. These are of the order of $10^{-9}$--$10^{-8}$C/m, producing, around dislocations or defects, local polarizations of several  $\mu$C/cm$^2$ and higher at low temperatures. The analysis of the behaviour of the flexoelectric current as a function of static bending in the low-temperature phase also suggests that the domain walls of STO are polarized, either intrinsically due to local gradient coupling, or extrinsically through defect accumulation. 
  
The authors thank Prof. S. A. T. Redfern for his experimental collaboration, Dr. M. Vopsaroiu at the National Physical Laboratory for help with piezoelectricity measurements, Prof. P. Thomas for SHG measurements, Prof. E. Artacho and Dr. M. Daraktchiev for valuable discussions, and acknowledge funding from Cambridge University  (PZ) and Marie Curie Fellowship (GC).


\begin{thebibliography}{500}
\bibitem{TagantsevReview} A. K. Tagantsev, Phase Transitions, \textbf{35}, 119 (1991).
\bibitem{CrossReview} L. E. Cross, J. Mat. Sci.,  \textbf{41}, 53 (2006).
\bibitem{Sharma} N. D. Sharma, R. Maranganti, P. Sharma, J. Mech. Phys. Solids in press (2007); R. Maranganti, N. D. Sharma, P. Sharma, Phys. Rev. B \textbf{74}, 014110 (2006).
\bibitem{MashkevichTolpygo} V.S. Mashkevich, K. B. Tolpygo, Zh. Eksp. eor. Fiz, \textbf{31}, 520 (1957) [Sov. Phys. JETP \textbf{4}, 455 (1957)].
\bibitem{Kogan}S. M. Kogan, Sov. Phys. Solid State, \textbf{5}, 2069 (1964).
\bibitem{TagantsevPRB} A. K. Tagantsev, Phys. Rev. B, \textbf{34}, 5883 (1986).
\bibitem{Ma_and_Cross} W.~Ma, L.~E.~Cross, Appl. Phys. Lett., \textbf{78}, 2920 (2001); 
\textbf{79} 4420 (2001);
\textbf{81} 3340 (2002);
\textbf{82} 3293 (2003);
\textbf{86} 072905 (2005).
\bibitem{Fousek} J. Fousek, L. E. Cross, D. B. Litvin, Material Letters \textbf{39}, 287 (1999)
\bibitem{Zhu_and_Cross} W. Zhu, J. Y. Fu, N. Li, L. E. Cross, Appl. Phys. Lett., \textbf{89}, 192904 (2006).
\bibitem{Catalan} G. Catalan, L. J. Sinnamon, J. M. Gregg, J. Phys.: Condens. Matter, \textbf{16}, 2253 (2004);  G. Catalan, B. Noheda, J. McAneney, L. J. Sinnamon,  J. M. Gregg,  Phys. Rev. B \textbf{72}, 020102(R) (2005).
\bibitem{Petzelt} J. Petzelt \emph{et al.}, Phys. Rev. B \textbf{64}, 184111 (2001)
\bibitem{Bednorz} J. G. Bednorz, K. A. M\"{u}ller, Phys. Rev. Lett., \textbf{52}, 2289 (1984).
\bibitem{Pertsev} N. A. Pertsev, A. K. Tagantsev, N. Setter, Phys. Rev. B, \textbf{61}, R825 (2000); \textbf{65}, 219901(E) (2002).
\bibitem{Haeni} J. H. Haeni \emph{et al.}, Nature \textbf{430}, 758 (2004)
\bibitem{Lin} C.-H. Lin,C.-M. Huang, G. Y. Guo, J. Appl. Phys., \textbf{100}, 084104 (2006).
\bibitem{Landau_Lifshitz}L. D. Landau, E. M. Lifshitz, \emph{Theory of Elasticity} (3rd ed. Pergamon, New York, 1986).
\bibitem{Buckley} A. Buckley, J. P. Rivera, E. K. H. Salje, J. Appl. Phys., \textbf{86}, 1653 (1999).
\bibitem{Kityk} A. V. Kityk, W. Schranz, P. Sondergeld, D. Havlik, E. K. H. Salje, J. F. Scott, Phys. Rev. B, \textbf{61}, 946 (2000).
\bibitem{Salje} B. Houchmandzadeh, J. Lajzerowicz and E. Salje, J. Phys. Cond. Mat. \textbf{3}, 5163 (1991). A. K. Tagantsev, E. Courtens and L. Arzel, Phys. Rev. B \textbf{64} 224107 (2001).
\bibitem{Vanderbilt} L. He and D. Vanderbilt, Phys. Rev. B \textbf{68}, 134103 (2003). 
\bibitem{Poindexter} E. Poindexter, A. A. Giardini, Phys. Rev. \textbf{110}, 1069 (1958).
\bibitem{rough} The magnitudes and signs of each of the calculated $f_{ijkl}$ depend on the absolute and relative magnitudes of all three of the measured $\mu$ values. Small errors in $\mu$ can lead to larger errors in  $f_{ijkl}$, thus the calculated values should be treated as order of magnitude estimates only.
\bibitem{dislocations} M.-W. Chu \emph{et al.}, Nature Mater. \textbf{3}, 87 (2004);\\ V. Nagarajan \emph{et al.}, Appl. Phys. Lett. \textbf{86}, 192910 (2005) 
\bibitem{Herger} R. Herger \emph{et al.}, Phys. Rev. Lett.  \textbf{98}, 076102 (2007)
\bibitem{Padilla_Vanderbilt} J. Padilla, D. Vanderbilt, Surface Science \textbf{418}  64 (1998)
\bibitem{Shirane} H. H\"unnefeld \emph{et al.}, Phys. Rev. B, \textbf{66}, 014113 (2002); S. Ravy \emph{et al.},  Phys. Rev. Lett. \textbf{98}, 105501 (2007)  and references therein.

\end{thebibliography}
\end{document}